# Gradient-induced Dzyaloshinskii-Moriya interaction

Jinghua Liang[1,4], Mairbek Chshiev[2], Albert Fert[3], Hongxin Yang[1,4,*]

[1] *National Laboratory of Solid State Microstructures, School of Physics, Collaborative Innovation Center of Advanced Microstructures, Nanjing University, Nanjing 210093, China*

[2] *Univ. Grenoble Alpes, CEA, CNRS, Spintec, 38000 Grenoble, France; Institut Universitaire de France, 75231 Paris, France*

[3] *Université Paris-Saclay, Unité Mixte de Physique CNRS-Thales, Palaiseau 91767, France*

[4] *Ningbo Institute of Materials Technology and Engineering, Chinese Academy of Sciences, Ningbo 315201, China*

*Email：hongxin.yang@nju.edu.cn

**ABSTRACT:** The Dzyaloshinskii-Moriya interaction (DMI) that arises in the magnetic systems with broken inversion symmetry plays an essential role in topological spintronics. Here, by means of atomistic spin calculations, we study an intriguing type of DMI (g-DMI) that emerges in the films with composition gradient. We show that both the strength and chirality of g-DMI can be controlled by the composition gradient even in the disordered system. The layer-resolved analysis of g-DMI unveils its additive nature inside the bulk layers and clarifies the linear thickness dependence of g-DMI observed in experiments. Furthermore, we demonstrate the g-DMI induced chiral magnetic structures, such as spin spirals and skyrmions, and the g-DMI driven field-free spin-orbit torque (SOT) switching, both of which are crucial towards practical device application. These results elucidate the underlying mechanisms of g-DMI and open up a new way to engineer the topological magnetic textures.

**KEYWORDS:** composition gradient, DMI, skyrmion, field-free SOT switching

Recently, the Dzyaloshinskii-Moriya interaction (DMI) has gained much attention since it is recognized to play an essential role in a broad range of physical phenomena, such as the observation of magnetic skyrmions [1,2,3], fast current-controlled domain wall motion [4,5], magnetoelectric effect [6,7], topological Hall effect [8], and magnetization switching [9,10]. The emergence of DMI requires the breaking of inversion symmetry in the spin system with spin-orbit coupling (SOC) [11,12]. In magnetic multilayers, it is usually believed that the disorders like the intermixing and dusting of elements at interfaces are detrimental to the interfacial DMI [13,14]. Interestingly, recent experiments [15,16,17,18] have observed a significant DMI (termed g-DMI hereafter) in the amorphous thin films with a composition gradient to break the inversion symmetry. The measured DMI value shows a linear relation with the film thickness, signifying a bulk nature of g-DMI despite in the thin films. In addition, several experiments [19,20,21,22,23] have suggested that the combination of g-DMI and spin-orbit torques (SOT) may lead to the current-induced magnetization switching in a single magnetic layer, although the microscopic mechanism is not well explored. To have an in-depth understanding of g-DMI, several fundamental questions need to be cleared up: How is the g-DMI controlled by the composition gradient? What about the distribution of g-DMI over the thin film? How the magnetic state and the spin dynamics get affected by the g-DMI?

In the present study, we present the systematic and comprehensive theoretical analysis of the g-DMI based on the atomistic spin calculations. We have also evidenced the g-DMI induced topological chiral magnetic structures and the g-DMI driven field-free SOT switching.

We consider a thin film consisting of $N = n_x \times n_y \times n_z$ cubic unit cells occupied by two kinds of atoms A and B with $n_x = n_y = 100$ used throughout this work. For our atomistic spin model, the A atoms act as magnetic transition element (e.g., Co or Fe), while the B atoms mainly play the role of heavy element with large SOC (e.g., Pt, Ta or Gd). For the system with linear vertical composition gradient, the composition $A_{x_i}B_{1-x_i}$ of the $i$th layer is characterized as the linear function $x_i = k(i - \frac{n_z+1}{2}) + \bar{x}$ with $|k| \leq \frac{1}{n_z}$ and $\frac{n_z|k|}{2} \leq \bar{x} \leq 1 - \frac{n_z|k|}{2}$ (see Figure S1 in the Supporting Information (SI)). Here $k$ and $\bar{x}$ are the gradient and the average of composition of the A atoms, respectively. In our simulations, we first generate a random configuration of the atoms according to the given $k$ and $\bar{x}$, and then perform the calculations based on such realization of composition. The calculation results are averaged over more than 10 trials for each composition, unless stated otherwise.

We have employed the Monte Carlo (MC) simulations to obtain the ground state and furthermore, solved the Landau-Lifshitz-Gilbert (LLG) equation to determine the spin dynamics. The atomistic spin Hamiltonian of thin film reads as

$$H = -\sum_{<i,j>} J_{ij} \mathbf{S}_i \cdot \mathbf{S}_j - \sum_{<i,j>} \mathbf{d}_{ij} \cdot (\mathbf{S}_i \times \mathbf{S}_j) - \sum_i K_i (S_z)_i^2 \quad (1)$$

with $i$ and $j$ being the A or B atom, and $\mathbf{S}_i$ and $K_i$ indicate the unit spin moment vector and the anisotropy parameter at site $i$, respectively. $J_{ij}$ and $\mathbf{d}_{ij}$ represent respectively the exchange constant and DMI vector for the nearest-neighbor atom pairs $<i,j>$. For the magnetic metallic system, the microscopic nature of $\mathbf{d}_{ij}$ can be well described by the Fert-Levy model[24] (see the calculation of the individual DMI vector by the Fert-Levy model in SI for details), stating that the DMI between that the two magnetic A atoms $i$ and $j$ can be mediated by the surrounding heavy B atoms at sites $l$. To make our simulations more realistic, the magnetic parameters are chosen based on the typical values of the Co-Pt thin film with perpendicular magnetic anisotropy (see the material parameters in SI).

We now present the properties of g-DMI. In the disordered thin films, the individual DMI vectors $\mathbf{d}_{ij}$ are randomly oriented, resulting in the vanishing total DMI. However, once the composition gradient is introduced, one should note that the calculated $\mathbf{d}_{ij}$ prefers the in-plane direction perpendicular to the bond between two A atoms in the same layer, and the out-of-plane components are almost cancelled out similar to the case of interfacial DMI[13,25,26]. Since the study of spin spiral is powerful way to describe and understand the underlying mechanism of DMI[13,27,28,29], we have quantified the effective g-DMI by the average DMI energy of the spin spiral

$$\langle \varepsilon_{DM} \rangle = -\frac{1}{N} \sum_{<i,j>} \mathbf{d}_{ij} \cdot (\mathbf{S}_i \times \mathbf{S}_j) \quad (2)$$

where $\mathbf{S}_i = [\cos(\mathbf{q} \cdot \mathbf{r}_i), 0, \sin(\mathbf{q} \cdot \mathbf{r}_i)]$ for the atom $i$ at the position $\mathbf{r}_i$. Here the long-wavelength spin spiral with wave vector $\mathbf{q} = \frac{2\pi}{a}(\frac{1}{100}, 0, 0)$ with $a$ being the lattice constant is adopted.

Figure 1(a) depicts the calculated dependence of the normalized $\langle \varepsilon_{DM} \rangle$ as a function of $k$

and $\bar{x}$ for $n_z = 15$. Clearly, the result shows that the g-DMI emerges only when $k \neq 0$. Moreover, the g-DMI has opposite sign for $k > 0$ and $k < 0$, and becomes stronger for larger $|k|$ with the maximums $|\langle \varepsilon_{DM} \rangle|_{max}$ located at $k = \pm \frac{1}{n_z}$ and $\bar{x} = 0.5$. These results imply an intuitive physical picture for the g-DMI, that is, the reversal of the gradient order changes the chirality of DMI and higher degree of inversion symmetry breaking leads to the stronger DMI. Moreover, we have examined the relationship between $\langle \varepsilon_{DM} \rangle$ and $k$ or $\bar{x}$ for different $n_z$, as shown in Figures 1(b) and 1(c). For given $\bar{x}$, $\langle \varepsilon_{DM} \rangle$ depends almost linearly with $k$ showing a minor change for different $n_z$ (see Fig. 1(b)), which indicates that the total DMI has a nearly linear correlation with both the composition gradient and film thickness. Such notable features have been identified in the GdFeCo[15,16] and CoPt[17] thin films. As discussed below, these are resulted from the additive nature of g-DMI inside the bulk of films. However, as $|k|$ can vary only between 0 and $\frac{1}{n_z}$, the larger range of $|k|$ and the resultant $|\langle \varepsilon_{DM} \rangle|_{max}$ can be obtained for smaller thickness $n_z$. For a given $k$, $\langle \varepsilon_{DM} \rangle$ can be enhanced by increasing $\bar{x}$ (see Fig. 1(c)), which mainly results from the increased number of the A atom pairs that contributes to the g-DMI.

To have a better understanding of the thickness dependence of g-DMI, it is necessary to know how the g-DMI distributes over the system. Figure 2 shows the calculated layer-resolved $\langle \varepsilon_{DM} \rangle$ for $k = 0$ and $\pm 0.01$ with $\bar{x} = 0.8$ and $n_z = 15$. One can see that in the absence of compositional gradient the interfacial DMI (gray bars in Fig. 2) is located at several layers near the top and bottom surfaces. First-principles calculations[13,25,26] on the interfacial DMI of different magnetic heterostructures have shown that the decay length of interface DMI is typically only a few atomic layers thick and determined by the constituent elements and the ordering of the atoms. Without composition gradient, the interfacial DMI appears due to the interfaces with the vacuum but is cancelled completely by the opposite DMI in the layers close to the opposite interface, and there is no net DMI inside the bulk part of the layer (layers $i = 5 \sim 11$). Accordingly, the total DMI vanishes. When the compositional gradient is introduced, the interfacial DMI is still opposite but becomes asymmetric at the opposite surfaces and, more notably, certain DMI emerges inside the bulk layers (see the red and blue bars in Fig. 2). We should note that, for each bulk layer and for a given sign of the gradient, $\langle \varepsilon_{DM} \rangle$ has the same sign and almost identical strength, indicating its additive nature. In other words, inside the thin film, the linear composition gradient results in an almost uniform distribution of g-DMI rather than the commonly expected gradient distribution[30]. Therefore, for a fixed $k$, the total g-DMI exhibits linear thickness dependence for the film that is not too thin to be comparable with the decay length of the interfacial DMI. One can also see that $\langle \varepsilon_{DM} \rangle$ of the bulk layers changes its sign with the reversal of compositional gradient. The g-DMI thus originates from the asymmetry distribution of composition along the thickness direction and spreads nearly homogeneously inside the bulk layers.

One of the prominent effects of DMI on the magnetic texture is the formation of chiral topological magnetic structures, e.g., spin spiral and skyrmion[1,2,31]. The topological protection of these stable spin configurations not only makes them attractive candidates for future high-density and low-power spintronic devices, but also triggers many fascinating phenomena such as the topological electric and spin Hall effects[8,32]. We now show that these topological structures can be induced by the g-DMI.

We have determined the magnetic ground state by MC simulation with periodic and open boundary conditions for in-plane and out-of-plane directions, respectively. By the fine tuning of $k$ that leads to the effective control of g-DMI, we have found rich topological chiral magnetic structures. For thin film with $\bar{x} = 0.8$ and $n_z = 15$, the ferromagnetically polarized state is preferred when $k \lesssim 0.019$, while the system can host a ground state of spin spiral with Néel-type domain walls when $k > 0.019$ (see Fig. 3(a) for $k = 0.026$), which is due to the increased g-DMI. Importantly, when $\bar{x}$ is changed to 0.5 so that the system has a larger tunable range of $k$, we find that the isolated skyrmion that is useful in the spintronic device application can be stabilized in the range $0.040 \lesssim k \lesssim 0.046$ (see Fig. 3(b) for $k = 0.044$). Interestingly, such skyrmion state can be identified as a skyrmion tube[33]. For $k > 0.046$, the skyrmion becomes unstable and the maze-like domains appear (see Fig. 3(c) for $k = 0.060$). The stabilization of different topological magnetic structures can be easily understood as a consequence of the competition between the g-DMI and the other magnetic interactions described in Eq. (1).

Here we should mention that, with the demonstration of the g-DMI induced topological magnetic textures, the resultant topological Hall effect can readily be expected. This has led us to presume that controversial hump-like Hall anomalies in the SrRuO$_3$ thin film, which were argued to be caused by either the interfacial DMI promoted topological Hall effect[34,35,36,37] or the inhomogeneous anomalous Hall effect[38,39,40], may be due to the g-DMI induced topological Hall effect resulted from the unintentionally introduced composition gradient during the process of fabrication.

In addition to magnetic structures, the DMI also has strong impact on the spin dynamics. The perpendicular magnetization switching by current-induced SOT is of fundamental interest as well as of practical relevance for the development of fast, non-volatile, and low-power consuming spintronic devices[9,10]. However, such SOT device typically requires an externally applied in-plane field to break the switching symmetry[41,42], which is an obstacle for practical application. Here, we will demonstrate that the field-free SOT switching can be achieved with the help of g-DMI.

In a single magnetic layer, a self-torque can be induced due to the unbalanced spin transfers between different parts of the sample. In this work, we have focused on the damping-like torque generated by spin-transfer from spin currents of spin Hall effect symmetry (see the discussion of SOT in the single magnetic thin layer in SI). To study the SOT switching probability distribution, we proceed as follows. First, the system is relaxed from the FM polarized state for 100 ps to obtain the relaxed state. After that, a charge current of 500 ps duration is applied along the +$x$-axis (the spin polarization direction $\bm{p} \propto \hat{\bm{z}} \times \bm{j}_c$ is then directed along +$y$-axis). Finally, the current is removed and the system is allowed to relax for another 100 ps to reach the final state. The switching probability is defined as $P = \frac{\langle S_z^r \rangle - \langle S_z^f \rangle}{2\langle S_z^r \rangle}$, where $\langle S_z^r \rangle$ and $\langle S_z^f \rangle$ is the average out-of-plane spin moment for the relaxed and the final state, respectively. In the following discussions, open boundary conditions in both the in-plane and out-of-plane directions, and $\bar{x} = 0.8$ and $n_z = 15$ are employed.

Figure 4 shows the calculated switching probability $P$ as a function of $k$ for different current density $j_c$. The non-zero $P$ indicates the onset of the magnetization switching. One can find that, with the increase of $k$, the switching probability $P$ first rises monotonously to a peak value and then falls steadily until it saturates at about 50%. For the considered conditions, the maximal $P$ that reaches nearly 100% occurs at $k = 0.019$ and $j_c = 0.88 \times 10^{13}$ A/m$^2$.

Since the system prefers the spin spiral state for large g-DMI as discussed in the previous section, $P$ tends to be 50% for the further increased $k$. The results manifest that the field-free SOT magnetization switching can be achieved when $k$ is larger than a critical value. Moreover, we should note that the critical $k$ is smaller for the higher $j_c$. Equivalently, the critical switching current density can be reduced by increasing the composition gradient. It should be noted that the magnitude of the critical switching current is affected by both the distribution of composition and the magnetic properties of the constituent elements. One can see from Fig. 4 that the critical switching current for the thin film with $k = 0.017$ is $0.86 \times 10^{13}$ A/m$^2$. This value is close to those calculated for the Co/Pt heterostructures ($0.26 \times 10^{13}$ A/m$^2$ in Ref.[43] and $\sim 1.50 \times 10^{13}$ A/m$^2$ in Ref.[44]) but is larger than that reported for Co-Tb thin film with composition gradient ($\sim 0.9 \times 10^{11}$ A/m$^2$ in Ref.[21]). As mentioned before, increasing the composition gradient can reduce the critical switching current of the Co-Pt thin film. Furthermore, the smaller effective exchange coupling constant in the ferrimagnetic Co-Tb thin film compared to that of the ferromagnetic Co-Pt can also lower the critical switching current.

To investigate the SOT switching mechanism, we have plotted the typical time evolution of $\langle S_z \rangle$ in Figure 5(a) for the aforementioned system with the maximal $P$. As explained below, we find that the g-DMI driven SOT switching is governed by the domain wall (DW) nucleation assisted by the edge spin canting and the subsequent DW propagation, similar to the case with interfacial DMI[43,44,45]. Prior to the injection of current, the relaxed magnetic structure exhibits edge spin canting with fixed chirality against the central part due to the appearance of g-DMI (see red arrows in the inset of Fig. 5(a)). When the current is applied, the damping-like SOT acting as a spin dependent effective field $\boldsymbol{B}_{sot} \propto (\boldsymbol{S} \times \boldsymbol{p})$ emerges. Owing to the spatial variation of the spin, $\boldsymbol{B}_{sot}$ tends to rotate the spin moments towards the film plane on one side while away from it on the opposite side, as schematically illustrated by the blue arrows in the inset of Fig. 5(a). As a result, a DW is nucleated on the edge where $\boldsymbol{B}_{sot}$ turns the spins towards the film plane. The nucleated DW is then moved to the opposite edge by the current[5] in the presence of g-DMI until it is expelled out of the film (see Figure 5(b) for the magnetic structures at several typical times), completing the magnetization switching.

In summary, we have systematically investigated the underlying mechanism of g-DMI. We have found that, in contrast to the interfacial DMI, the g-DMI can be engineered by the composition gradient without the demand of high-quality interface for the enhancement of total DMI. We have found that the linear composition gradient can lead to almost uniform distribution of g-DMI inside the bulk layers of thin film. Due to the additive nature of g-DMI, increasing the film thickness is beneficial to increase the total DMI. Moreover, we have evidenced the g-DMI induced topological magnetic structure and the g-DMI driven field-free SOT switching, both of which are crucial towards the practical spintronic device applications. Compared to the interface DMI with its properties determined by constituent elements and ordering of the atoms at the interface, both the strength and chirality of g-DMI is continuously tunable by the composition gradient, being thus advantageous for practical applications. Since the g-DMI can stabilize many different types of topological structures and has strong impact on the spin dynamics, one can envision the discovery of many more exotic spin phenomena induced by the g-DMI in the future.

**Methods.** The MC simulation is performed using the Metropolis algorithm. In the simulations, we have gradually cooled the system down from 1000K to 10K in the step of 10K. For each temperature, $10^5$ MC steps are employed to thermalize the system.

The spin dynamics are studied by solving the LLG equation augmented with the SOT term:

$$\frac{\partial \boldsymbol{S_i}}{\partial t} = -\gamma \boldsymbol{S_i} \times \boldsymbol{B}_{eff} + \alpha \boldsymbol{S_i} \times \frac{\partial \boldsymbol{S_i}}{\partial t} + \boldsymbol{T}, \quad (3)$$

where $\boldsymbol{B}_{eff} = -\frac{\delta H}{\mu_i \delta \boldsymbol{S_i}}$ is effective magnetic field acting on the atom $i$, with $H$ being the spin Hamiltonian given by Eq. (1) in the main text and $\mu_i$ being the spin moment. $\gamma$ is the gyromagnetic ratio, $\alpha$ is the Gilbert damping parameter. $\boldsymbol{T}$ is the damping-like SOT. The LLG equation is solved using the Heun's method with a time step of 0.1 $f$s.

**Supporting Information.** Plot of the linear thickness dependence of the composition of the A atom; calculation of the individual DMI vector by the Fert-Levy model; discussion of SOT in the single magnetic thin layer; material parameters; estimation of the effective DMI parameter for g-DMI.

**Acknowledgments.** This work was supported by the National Key Research and Development Program of China (MOST) (Grant No. 2022YFA1405102), the National Natural Science Foundation of China (Grant Nos. 11874059 and 12174405), the Key Research Program of Frontier Sciences, CAS (Grant No. ZDBS-LY-7021), Ningbo Key Scientific and Technological Project (Grant No. 2021000215), "Pioneer" and "Leading Goose" R&D Program of Zhejiang Province (Grant No. 2022C01053), Zhejiang Provincial Natural Science Foundation (Grant No. LR19A040002), Beijing National Laboratory for Condensed Matter Physics (Grant No. 2021000123), European Union's Horizon 2020 research and innovation Programme under grant agreement 881603 (Graphene Flagship).

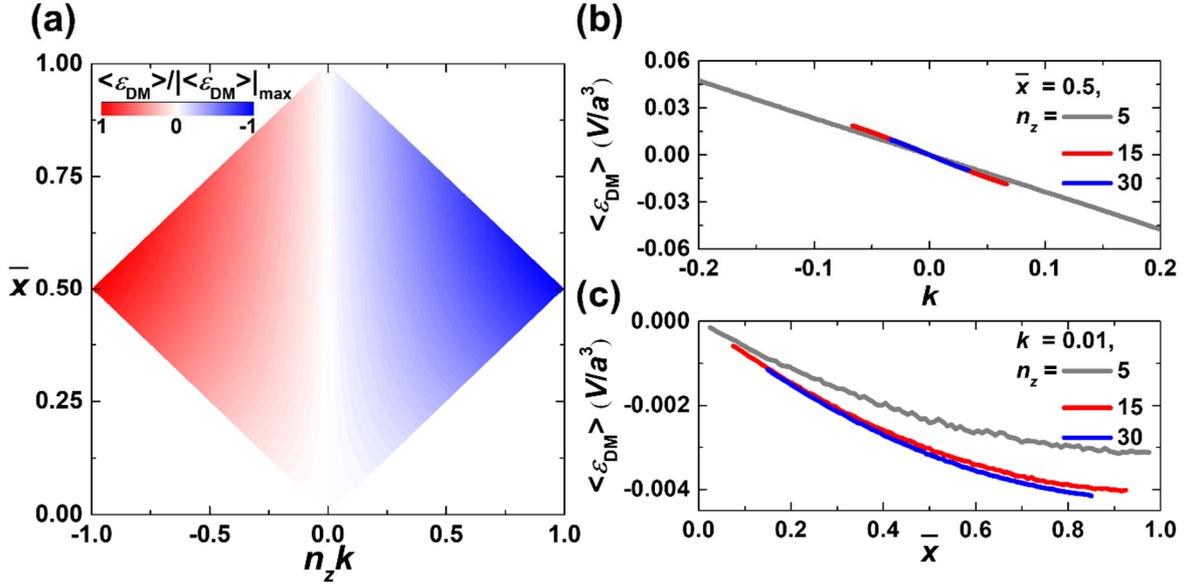

**Figure 1.** (a) The calculated of $\langle \varepsilon_{DM} \rangle / |\langle \varepsilon_{DM} \rangle|_{max}$ as a function of $k$ and $\bar{x}$ for $n_z = 15$. $|\langle \varepsilon_{DM} \rangle|_{max}$ is the maximum of $\langle \varepsilon_{DM} \rangle$ that locates at $k = \pm \frac{1}{n_z}$ and $\bar{x} = 0.5$. (b) and (c) are the calculated $\langle \varepsilon_{DM} \rangle$ as a function of $k$ and $\bar{x}$, respectively, for different $n_z$. The unit of $\langle \varepsilon_{DM} \rangle$ is given in $V/a^3$ with $V$ being the parameter quantifying the DMI strength provided by the B atom.

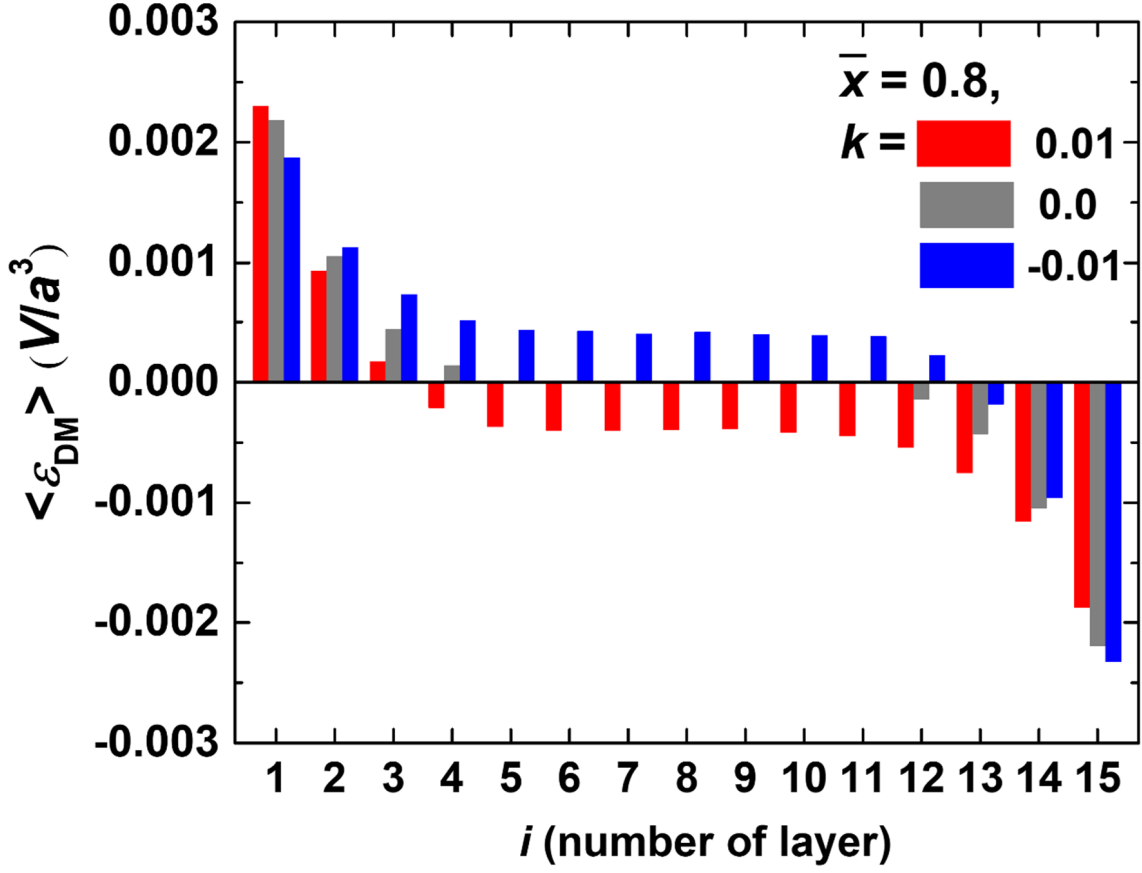

**Figure 2.** Layer-resolved $\langle \varepsilon_{DM} \rangle$ for thin films of $k = 0$ (grey bars), 0.01 (red bars) and -0.01 (blue bars) with $\bar{x} = 0.8$ and $n_z = 15$. The total averaged DMI energies obtained by summing the layer-resolved $\langle \varepsilon_{DM} \rangle$ of all layers for $k = 0.01$, 0, and -0.01 are -0.004, 0, and 0.004 ($V/a^3$), respectively.

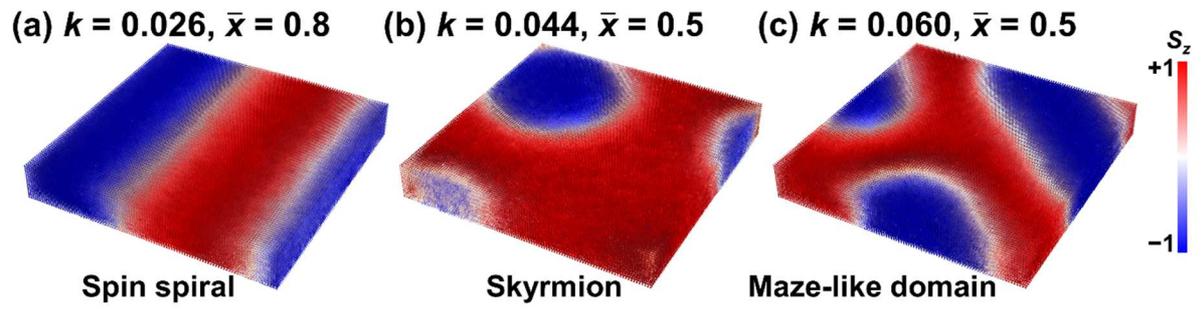

**Figure 3.** The chiral magnetic textures including (a) spin spiral, (b) skyrmion, and (c) maze-like domain obtained by the MC simulations for thin films with different $k$ and $\bar{x}$. The arrows indicate the direction of the spin moments with the out-of-plane component $S_z$ encoded by the color map.

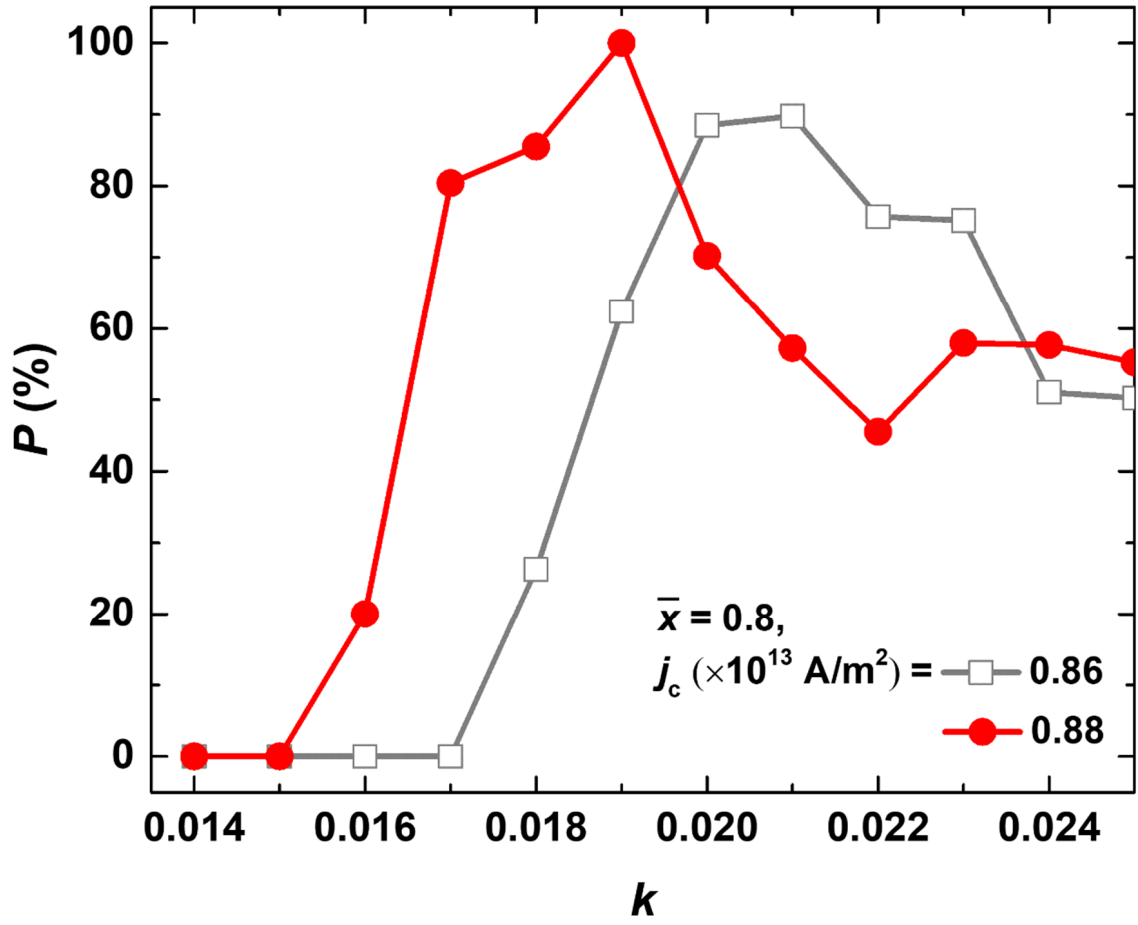

**Figure 4.** The calculated switching probability $P$ as a function of $k$ for $\bar{x}$ =0.8, and $j_c$ = $0.86\times10^{13}$ A/m$^2$ (open squares) and $0.88\times10^{13}$ A/m$^2$ (solid dots).

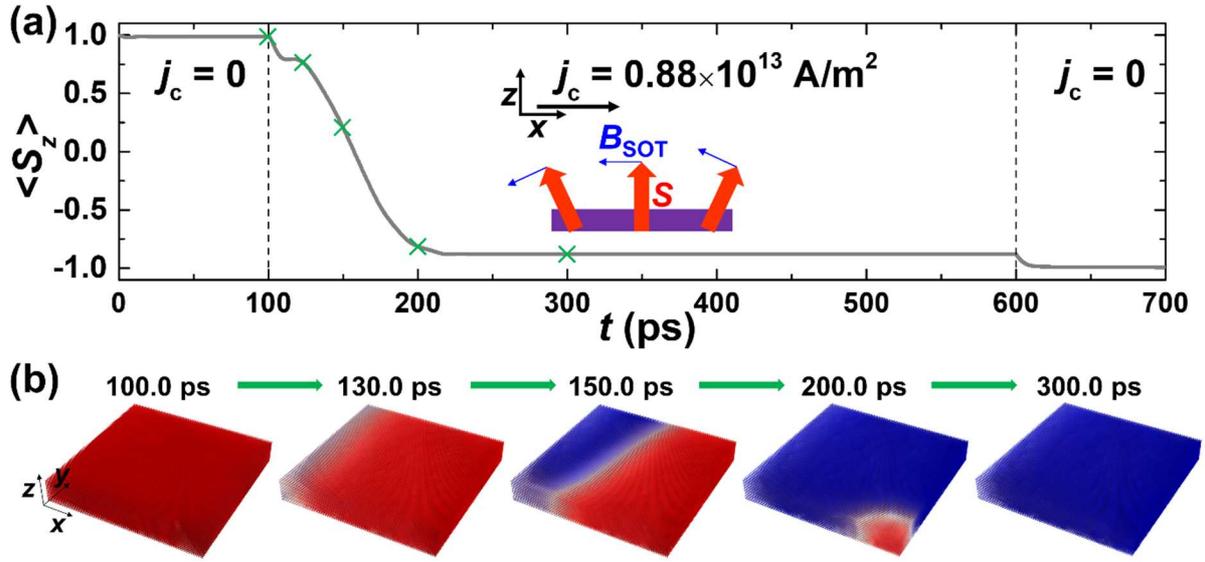

**Figure 5.** (a) The typical time evolution of $\langle S_z \rangle$ for the thin film of $k = 0.019$ and $\bar{x} = 0.8$. The current of $j_c = 0.88 \times 10^{13}$ A/m² is applied from 100ps to 600ps. The inset schematically illustrates the spatially dependent spin canting (red arrows) and effective SOT field (blue arrows) at 100ps. (b) The magnetic structures corresponding to the times marked with green × symbol in (a). The arrow and color codes are the same as Figure 3.

# Supporting Information for
# Gradient-induced Dzyaloshinskii-Moriya interaction

Jinghua Liang[1,4], Mairbek Chshiev[2], Albert Fert[3], Hongxin Yang[1,4,*]

[1] *National Laboratory of Solid State Microstructures, School of Physics, Collaborative Innovation Center of Advanced Microstructures, Nanjing University, Nanjing 210093, China*

[2] *Univ. Grenoble Alpes, CEA, CNRS, Spintec, 38000 Grenoble, France; Institut Universitaire de France, 75231 Paris, France*

[3] *Université Paris-Saclay, Unité Mixte de Physique CNRS-Thales, Palaiseau 91767, France*

[4] *Ningbo Institute of Materials Technology and Engineering, Chinese Academy of Sciences, Ningbo 315201, China*

*Email：hongxin.yang@nju.edu.cn

**(S1) Plot of the linear thickness dependence of the composition of the A atom**

In the main text, we consider the thin film with linear vertical composition gradient. Figure S1 show the linear thickness dependence of the composition of the A atom, where $x_i = k(i - \frac{n_z+1}{2}) + \bar{x}$ is the composition of the A atom at the $i$th layer.

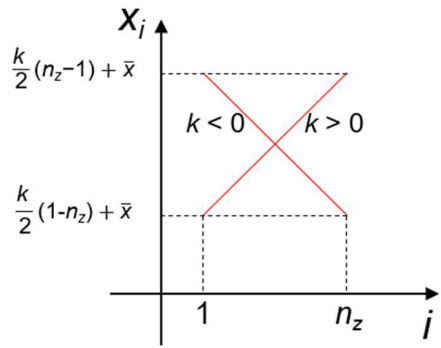

**Figure S1.** The schematical plot of the composition of the A atom as a linear function of layer number.

## (S2) Calculation of the individual DMI vector by the Fert-Levy model

For the magnetic metallic system, the microscopic nature of DMI can be well described by the Fert-Levy model[45], which indicates that the DMI between that the two magnetic A atoms $i$ and $j$ can be mediated by the heavy B atoms at sites $l$ with the DMI vector $\boldsymbol{d}_{ij}$ satisfies

$$\boldsymbol{d}_{ij} = \sum_l \boldsymbol{d}_{ij-l} = V \sum_l \frac{(\boldsymbol{R}_{li} \cdot \boldsymbol{R}_{lj})(\boldsymbol{R}_{li} \times \boldsymbol{R}_{lj})}{|\boldsymbol{R}_{li}|^3 \cdot |\boldsymbol{R}_{lj}|^3 \cdot |\boldsymbol{R}_{ij}|} \quad (S1)$$

where $\boldsymbol{R}_{li}$, $\boldsymbol{R}_{lj}$, and $\boldsymbol{R}_{ij}$ are the corresponding distance vectors and the summation is over the surrounding B atoms. The parameter $V$ defining the DMI strength is a material specific quantity related to the SOC of the heavy atom B. Note that in our model only the A atoms pairs have a nonvanishing DMI vectors $\boldsymbol{d}_{AA-B}$ mediated by the B atoms, while the others ($\boldsymbol{d}_{AA-A}$, $\boldsymbol{d}_{AB-A}$, $\boldsymbol{d}_{AB-B}$, $\boldsymbol{d}_{BB-A}$, and $\boldsymbol{d}_{BB-B}$) are zero. As an example, when two A atoms and a B atom form a triangle on a surface of cubic unit cell with lattice constant $a$, as shown in Figure S2, the calculated DMI vector $\boldsymbol{d}_{AA-B}$ is along the $+x$ direction and $|\boldsymbol{d}_{AA-B}| = \frac{\sqrt{2}}{4} \frac{V}{a^3}$. In our calculations, a cutoff of $5a$ is used for the distances involved with the B atoms in Eq. (S1).

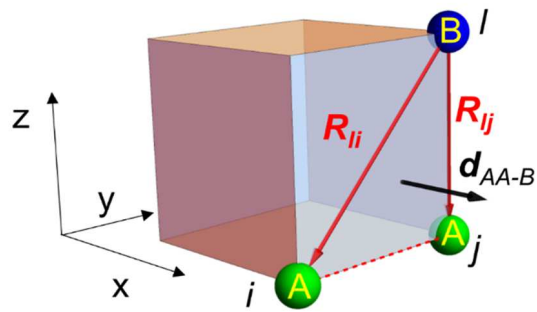

**Figure S2.** For two A atoms (green balls) and one B atom (blue ball) that form a triangle on the surface of a cubic unit cell, the DMI vector $\boldsymbol{d}_{AA-B}$ calculated by the Eq. (S1) is along the $+x$ direction and $|\boldsymbol{d}_{AA-B}| = \frac{\sqrt{2}}{4} \frac{V}{a^3}$.

## (S3) Discussion of SOT in the single magnetic thin layer

In principle, a SOT term $T$ is due to the spin transfer from a spin current induced by SOC into the magnetization of a magnetic layer[45]. For example, SOT can be due to the transfer of a spin current induced by spin Hall effect (SHE) in Pt from the Pt layer into a magnetic layer. In an isolated single magnetic layer, SOC-induced spin currents of both SHE and spin anomalous Hall effect (SAHE) symmetries exist, for example, when a spin current is created in a part with large SOC and transferred into a part with smaller SOC, but the global transfer is zero in most cases as the spin transfer from the magnetization to the spin current is compensated by the spin transfer of the spin current into the magnetization in another part of the system. However, unbalanced non-zero internal spin transfer and resulting SOT in single magnetic layer can subsist, for example, when a part of the spin current can be transferred outside in another layer or if there is an asymmetry between the additional spin-lattice relaxation in the respective parts of the layer with outgoing and incoming transfers[45,45]. For our studied single magnetic thin film with a composition gradient, the SHE-like spin current can generate a damping like torque on the magnetization due the imbalance between the transfer to the bottom magnetization of spin currents emitted from the top layers and the transfer to the top magnetization of spin currents with opposite polarization emitted from the bottom layers, while the SAHE term that polarizes along with the magnetization cannot generate a torque on magnetization[Error! Bookmark not defined.]. Therefore, we have focused on the damping-like SOT $T = -\beta_{DL} S_i \times (S_i \times p)$ generated via the SHE in the present work. Here $\beta_{DL} = \frac{\gamma \hbar J_c \theta_{sh} a^2}{2 e \mu_i}$ and $p$ is the direction of the spin polarization. $J_c$ is the value of the charge current density $J_c$, $\theta_{sh}$ is the spin Hall angle, $\hbar$ is the reduced Planck constant, and $e$ is the absolute value of electron charge.

## (S4) Material parameters

In the MC simulation and the study of spin dynamics, we have chosen the material parameters based on the Co-Pt thin film. The typical material parameters for the Co thin film are [45]: exchange coupling $J = 1.27 \times 10^{-20}$J, magnetic anisotropy $K = 3.14 \times 10^{-23}$J, spin moment $\mu = 3.03 \times 10^{-23}$J/T. Moreover, from our previous research[45], the typical DMI constant for the Co/Pt interface is $d = 6.41 \times 10^{-22}$J. If we assume that $|d_{AA-B}| = \frac{\sqrt{2}}{4} \frac{V}{a^3} = d$, where $d_{AA-B}$ is the DMI vector shown in Fig. S2, it can be deduced that $V/a^3 \approx 0.14J$. We have

set the material parameters as in the following table.

| Material Parameter | Value |
|---|---|
| exchange coupling constant between the two nearest neighboring A atoms $J_{A-A}$ | $J$ |
| exchange coupling constant between the nearest neighboring A and B atoms $J_{A-B}$ | $0.1J$ |
| exchange coupling constant between the two nearest neighboring B atoms $J_{B-B}$ | $0.1J$ |
| DMI strength parameter $V/a^3$ | $0.14J$ |
| anisotropy parameter of A atom $K_A$ | $K$ |
| anisotropy parameter of B atom $K_B$ | $K$ |
| spin moment of A atom $\mu_A$ | $\mu$ |
| spin moment of B atom $\mu_B$ | $\mu$ |
| lattice constant $a$ | 3Å |
| Gilbert damping parameter $\alpha$ | 0.31 |
| spin Hall angle $\theta_{sh}$ | 0.1 |

**Table S1.** The chosen material parameters for the calculations.

**(S5) Estimation of the effective DMI parameter for g-DMI.**

As in micromagnetic simulations and experiments the micromagnetic DMI parameter $D$ is often given in units of J/m², one can estimate the effective DMI parameter $D$ of g-DMI based on the calculated DMI energy density. In the long wavelength limit, the average DMI energy $\langle \varepsilon_{DM} \rangle$ per unit cell scales linearly with the length of the spin spiral wave vector[45,45,45], i.e., $\frac{\langle \varepsilon_{DM} \rangle}{a^3} = Dq$, where $a^3$ is the volume of unit cell. One can thus estimate the micromagnetic DMI parameter as $D = \frac{\langle \varepsilon_{DM} \rangle}{qa^3}$. Using $q = \frac{2\pi}{100a}$ and based on the materials parameters given in S4, the calculated micromagnetic DMI parameter $D$ corresponding to the cases in Fig. 1(b) are shown in the following Figure S3.

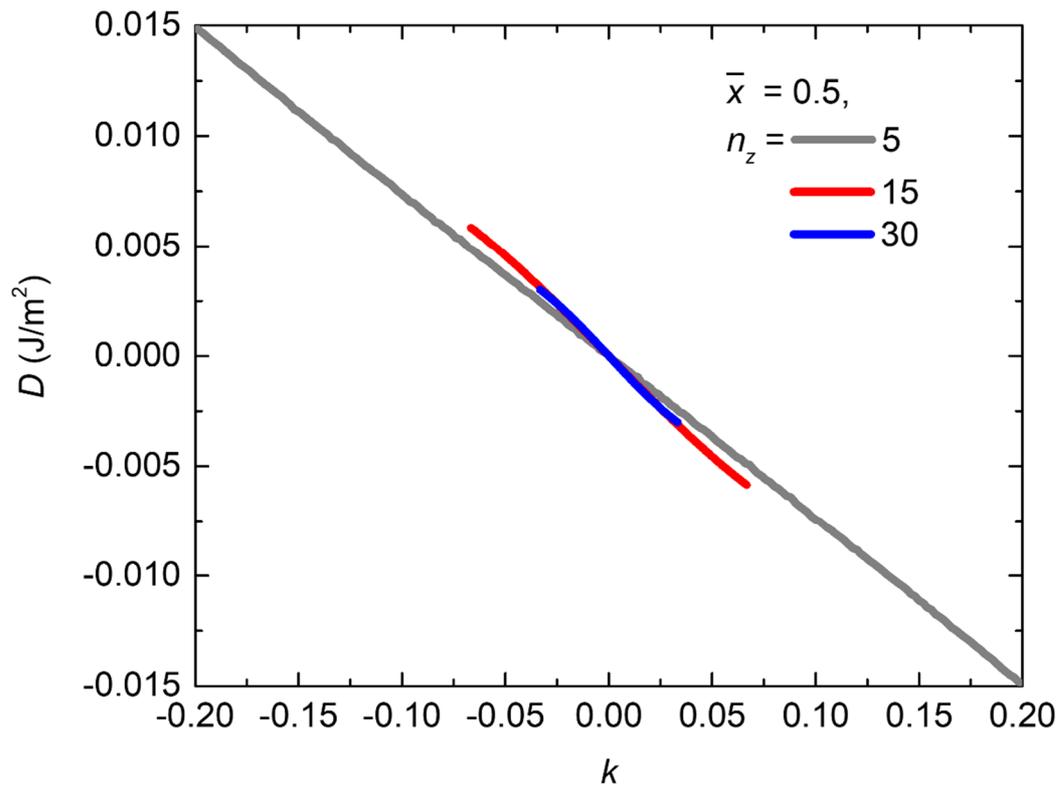

**Figure S3.** The calculated effective micromagnetic DMI parameter $D$ corresponding to the cases in Fig. 1(b).